\documentclass[prl,aps,preprint,showpacs]{revtex4}
\usepackage{graphicx}
\usepackage{times}
\usepackage{dcolumn}
\usepackage{natbib}

\newcommand{\EE}{{\cal E}}

\begin{document}

\title{Exact description of self-focusing in highly nonlinear geometrical optics}
\author{Larisa L. Tatarinova}
\author{Martin E. Garcia}
\affiliation{Theoretische Physik, FB 18, Universit\"at Kassel,
Heinrich-Plett-Str.\ 40, 34132 Kassel, Germany}
\date{\today}

\begin{abstract}
We demonstrate that laser beam collapse in highly nonlinear media can be described,
 for a large number of experimental conditions, by the geometrical optics 
approximation within high accuracy.  Taking into account this fact we succeed in 
constructing analytical 
solutions of the eikonal equation,  which are exact on the beam axis and provide:
 i) a first-principles determination of the self-focusing position, thus replacing 
the widely used empirical Marburger formula, ii) a benchmark solution for numerical
 simulations, and iii) a tool for the experimental determination of the high-order 
nonlinear susceptibility. Successful comparison with several experiments is presented.
\end{abstract}

\maketitle

Nonlinear light self-focusing is a self-induced modification of the
optical properties of a material which leads to beam collapse at a
certain point $z_{\rm sf}$ in the media. This effect, first
observed in the 1960s, plays nowadays a key role in all
scientific and technologycal applications related to the propagation
of intense light beams \cite{Diels}, like material processing \cite{Englert}, environmental sciences \cite{Woeste},
femtochemistry in solutions \cite{Chem}, macromolecule
chromatography \cite{MacrMol}, medicine \cite{Med}, etc.  

Usually,
$z_{\rm sf}$ is estimated using the empirical Marburger formula
\cite{Diels,CPhysRep,BergeRPP}, which has been constructed via fitting
 the results of extensive numerical simulations obtained for the case
when the refractive index $n$ is a linear function of the electric
field intensity $n=n(I)=n_0+n_2I$, ($n_2>0$) \cite{Marburger,ShirKovPhysRep,Akhmanov,Prep}.
 In most modern experiments, however,  high beam intensities are used  
 for which the linear approximation breaks down, and
further contributions to $n(I)$ must be considered
\cite{BergePhysRep,CPhysRep,BergeRPP}. For these cases no general mathematical condition 
for the behavior of $z_{\rm sf}$ and the filament intensity has been 
 derived so far.  Most theoretical results  are  based on numerical studies, or on
variational calculations assuming a fixed beam profile inside 
 the medium (see e.g. \cite{CPhysRep} and
Refs. therein). An analytical theory, able to accurately describe beam
collapse in highly nonlinear optics, is
still missing. Moreover, it is widely believed that 
the exact treatment of beam propagation in a highly nonlinear medium
can only be done numerically \cite{Diels}.

In this letter, we construct for the first time analytical solutions
for the eikonal equations with highly nonlinear forms of the
refractive index avoiding any {\it a priori} assumptions on the form
of the beam during propagation. The results obtained are exact on the
beam axis within the geometrical optics approximation, 
which we demonstrate to be accurate for many of the situations taking place in modern 
experiments.  
Our approach permits not only to obtain exact expressions for $z_{\rm sf}$ for
different nonlinear functions $n(I)$, but also to find a general mathematical framework 
  which corrects traditionally used
formulas for the filament intensity calculation \cite{CPhysRep}. 
  Since
the accuracy of the semi-classical approximation can be easily
estimated, we can determine and control the error in our calculations, which
is not possible in the case of the Marburger formula. 

Based on these results we are also able to propose experiments to precisely determine the high 
order nonlinear susceptibility of different materials.
Moreover, our results yield a natural explanation of the experimentally observed
\cite{Fibich} deviation of the scaling law $z_{\rm sf}\sim I^{-1/2}$
for high beam intensities. 

We consider the propagation of a linearly polarized laser beam of
initially Gaussian shape.  Starting from the nonlinear wave equation
and assuming that the light beam is almost monochromatic and that the
envelope varies slowly in space and time, one obtains a generalized
nonlinear Schr\"odinger equation (NLSE) of the form \cite{Diels}:
\begin{eqnarray}
i\partial_z \EE+{1\over 2k_0}\partial_{xx}\EE+k_0n(\vert \EE\vert^2)\EE=0.
\label{NS}
\end{eqnarray}
where $\EE$ is the electric field, $z$ the propagation length and $k_0$
the wave vector. The second term describes wave diffraction on the transverse
plane. $n(\vert\EE\vert^2)$ is the nonlinear refractive index. The magnitude of the contributions of the diffraction and the nonlinear effects to 
the beam propagation can be estimated through the comparison of the characteristic distances $L_{\rm diff}$ and $L_{\rm nl}$, at which the beam suffers considerable changes \cite{CPhysRep}. Then, $L\equiv L_{\rm nl}/L_{\rm diff}$ is a messure of the accuracy of the geometrical optics approximation: if $L\ll 1$, diffraction can be neglected. 

\begin{figure}
\includegraphics[angle=0,width=10cm]{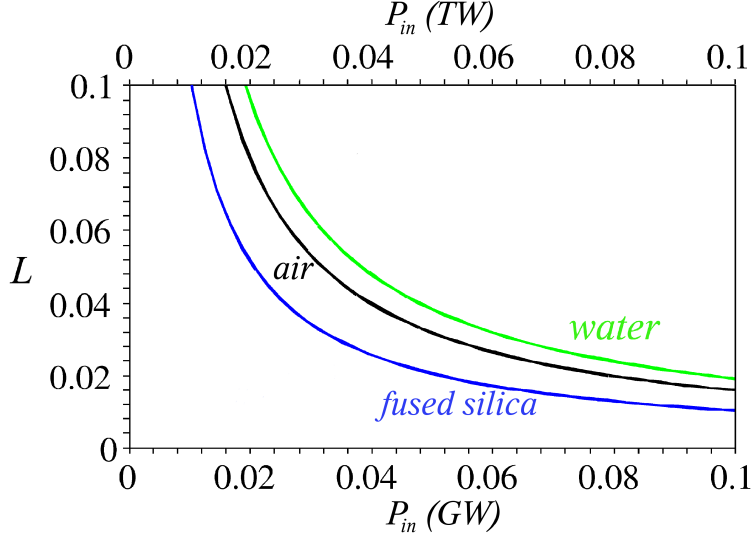}
\caption{\label{SemClass} (Color online.) Accuracy  $L\equiv L_{\rm nl}/L_{\rm diff}$ 
of the geometrical optics approximation for different media 
 as a function of the laser pulse power $P_{\rm in}$. The green, black and blue curves refer to 
water, air and fused silica, respectively. 
 The pulse wavelength is assumed to be $800{\rm nm}$.  $P_{\rm in}$ is given in units of ${\rm TW}$ for
 air, and of ${\rm GW}$ for  water and fused silica.}
\end{figure}

The main contribution to $n(\vert \EE \vert^2)$ is usually given by
the Kerr cubic term $n_2\vert \EE\vert^2$. Therefore it is natural 
to define a nonlinear length $L_{\rm nl}=1/(k_0 n_2 I_0)$, where $I_0$ is the intensity of the beam at the entry
plane of the nonlinear medium.  The diffraction length is deffined as $L_{\rm diff}=n_0k_0w_0^2/2$, where $w_0$ is the
initial beam radius \cite{CPhysRep}. In Fig.\ref{SemClass} we plot $L$ as a
function of the initial beam power $P_{\rm in}$ ($P_{\rm
in}=I_0w_0^2\pi/2$) for different media. In many recent
experiments $L\sim 0.05$ or smaller (see \cite{CPhysRep} and Refs. therein), thus making the
geometrical optics approximation valid.

We represent the electric field $\EE$ in Eq.~(\ref{NS}) in
the eikonal form: $\EE=\sqrt{I}\exp(ik_0S)$. Neglecting the
diffraction term we obtain
\begin{eqnarray}
\partial_z S=-{1\over 2}(\partial_x S)^2+n(I),\hspace{0.3cm}
\partial_z I= - \partial_x( I\partial_x S).
\label{eikonaleq}
\end{eqnarray}
As a next step we rewrite these equations in a dimensionless form by considering  
 $x$ and $z$ in units of $w_0$ and the
beam intensity $I$ in units of  $I_0$. We also introduce the dimensionless variable
$v\equiv \partial_x S$ and differentiate the first of Eqs.~(\ref{eikonaleq}) with respect to 
$x$. The refractive index term yields
$\partial_x n(I)\equiv a\varphi \partial_x I$, where
$\varphi = \varphi(I)\equiv \partial_I n(I)$ and $a\equiv n_2I_0$ is a dimensionless
parameter. The order of magnitude of $a$ in a large number of modern experiments lies below
$ 10^{-5}$ \cite{CPhysRep}. It is therefore reasonable to consider $a$ as a small parameter.

The boundary value problem to be solved can be summarized as 
\begin{eqnarray}
\partial_z v+v\partial_x v-a\varphi \partial_x I=0,\hspace{0.3cm}&&\partial_z I+v\partial_x I+I\partial_x v=0,\nonumber\\
v(0,x)=0,\hspace{2cm} &&I(0,x)=\exp (-x^2),\label{BE}
\end{eqnarray}
which describes the propagation of an initially collimated Gaussian beam in a
nonlinear medium. Solutions of Eqs.~(\ref{BE}) and their derivatives can exhibit singularities for
particular values of $z$. Analyzing these points, we obtain the
nonlinear self-focusing position $z_{\rm sf}$ of the laser beam \cite{noncollim}. 

Following  Ref.~\cite{ShirKovPhysRep}, one can notice that the system
(\ref{BE}) is linear with respect to the first order derivatives.
Therefore, it is convenient to use a hodograph transformation \cite{Bocharov} in order
to transform it into a linear system of the form
\begin{eqnarray}
\partial_w\tau-{I\over\varphi}\partial_I\chi=0,\hspace{0.3cm}
\partial_w\chi+a\partial_I\tau=0,\label{hod}
\end{eqnarray}
where $\tau=I z$, $\chi=x-v z$, $w=v/a$. The boundary conditions are
transformed as: for 
$w=0,$ $\chi=\sqrt{\ln(1/I)}$ and $\tau=0.$

Taking into account the smallness of $a$ we solve Eqs.~(\ref{hod}) by proceeding 
 in two steps \cite{Kuramoto,GKZ}. 
First, accounting for the boundary conditions, from the first of Eqs.~(\ref{hod}), we find $\tau$ as a function of $\chi,$ $I$ and $w$:
\begin{eqnarray}
\tau=-w/( 2\chi \varphi)\label{tau}.
\end{eqnarray}
Then, substituting Eq.~(\ref{tau}) into the second of Eqs.~(\ref{hod}),
 we obtain a closed partial
differential equation for the variable $\chi$
\begin{eqnarray}
\partial_w\chi+{aw\over 2\chi^2\varphi}\partial_I\chi+
{a w \over 2 \chi \varphi^2}\partial_I\varphi=0.\label{EQ}
\end{eqnarray}
Integration of Eq.~(\ref{EQ}) results in two invariants
\begin{eqnarray}
\chi\varphi=\Psi_1,\hspace{0.5cm}\int \varphi^{-1}dI-a\tau^2=\Psi_2. \label{sol}
\end{eqnarray}
Then, the final steps for the construction of the  desired solutions of Eqs.~(\ref{BE}) can be summarized  as  follows.
 With the help of Eqs.~(\ref{sol}) we express $I$ and $\chi$ as
 functions of the integration invariants: $I=I(\Psi_1,\Psi_2)$
and $\chi=\chi(\Psi_1,\Psi_2)$. Then, we require that according to the boundary conditions for $\tau=0$ the equation $I(\Psi_1,\Psi_2)=\exp[-\chi(\Psi_1,\Psi_2)^2]$ must be fulfilled. Finally, substituting $\Psi_1$ and $\Psi_2$ into the relation above and returning to the original variables, we get the solution of Eqs.~(\ref{BE}). 

 The scheme presented above allows us to find 
  analytical solutions of the optics equations for different
 types of nonlinearities $\varphi(I)$.  
 For high field intensity, the refractive index of most materials  
 contains nonlinear contributions  additional to the Kerr-term $n_2I$.
 Usually, they are modeled as a power function of the  
intensity in the general form $\beta I^K$.  Physically, this term can
be attributed to the fifth order nonlinear susceptibility $n_4I^2$
\cite{Berge_Chi5,CPRA} or to the material ionization $\sigma_K I^K$, where $K$
is the number of photons absorbed, and $\sigma_K$ being the multiphoton
ionization (MPI) cross section \cite{CPhysRep}.

In the following we use our approach to determine under which conditions self-focussing takes place or can be prevented by the high order nonlinearity, and arrive at a new general mathematical equation to obtain 
 the filament intensity, which improves previous theories. 
 Let us first consider a system having a nonlinear part of the refractive index of the form  $n(I)=n_2I-n_4 I^2$ (i.e., $K=2$). 
 In this case  $\varphi = 1-\beta I$, 
where $\beta\equiv 2n_4I_0/n_2$. Substituting $\varphi$ into Eqs.~(\ref{sol}), we get
$\Psi_2=-1/\beta \ln(1-\beta I)-a\tau^2$ and 
$\Psi_1=\chi(1-\beta I)$. 
Thus, the solution to Eqs.~(\ref{BE}) reads
\begin{eqnarray}
&&1-(1-\beta I)e^{aI^2z^2\beta}=\beta \exp\left({-x^2 e^{-2a\beta I^2z^2} \over [1-2aIz^2(1-\beta I)]^2}\right),\nonumber\\*
&&v=-2aIzx(1-\beta I)/[1-2aIz^2(1-\beta I)].\nonumber
\end{eqnarray}
After differentiating these expressions with respect to $x$ and $z$, solving the obtained
system of four algebraic equations with respect to $\partial_x I$,
$\partial_t v$ etc., and substituting the resulting expressions into
Eqs.~(\ref{BE}), one can verify that the obtained solutions  are {\em exact} on the beam axis ($x=0$, $v\vert_{x=0}=0$) \cite{Note}. 
The on-axial beam intensity distribution is given by
\begin{eqnarray}
aI^2 z^2 \beta = \ln \left[{1-\beta\over 1-\beta I}\right],\label{sol_(1-bn)}
\end{eqnarray}

Analyzing the Eq.~(\ref{sol_(1-bn)}) we find  a critical value
$\beta_c\sim 0.175$. For
$\beta_c > 0.175$, Eq.~(\ref{sol_(1-bn)}) has no special points, the on-axial intensity monotonically increases and
 reaches a saturation value $I_{\rm sat}$. 
 By analyzing the asymptotic behaviour of $I=I(z)$ we obtain $I_{\rm sat}=1/\beta$. 
 Note that this value fullfils 
 the condition  $1-\beta I_{\rm sat} = \varphi (I_{\rm sat}) = 0$. 
   For $\beta_c < 0.175$ there is an
interval $[z_1,z_2 ]$ on the beam axis where the solution $I(z)$ is not unique. The
first point $z_1$ can correspond to development of an instability
in the beam, while the second point $z_2$ corresponds to $z_{\rm sf}$, where the intensity sharply increases, and  
the beam is compressed into a single filament with $I_{\rm sat}$.

For materials described by $n(I)=n_2I-n_6 I^3$, we have $K=3$, 
$\varphi=1-\beta I^2$, $\beta =3n_6I_0^2/n_2$ and the  on-axial intensity
distribution is given by
\begin{eqnarray}
 {\rm arctanh}(\sqrt{\beta} I)-aI^2z^2\sqrt{\beta}=
{\rm arctanh}\sqrt{\beta}.\label{sol_(1-bn2)}
\end{eqnarray}
 From the analysis of the asymptotic behavior ($z \to \infty$) we obtain $\beta_c\sim 0.05,$
$I_{\rm sat}= 1/\sqrt{\beta}$. By inspection, we realize again that, as for $K=2$, 
 the intensity of the beam saturates when  $\varphi = 0$. 

Notice that the values of $I_{\rm sat}$ for $K=2$ and 3 obtained here are different
from previous theoretical estimates \cite{CPRA,CPhysRep,TatGar}, 
which were obtained assuming that
the intensity in the filament saturates when the nonlinear terms in $n(I)$ 
compensate each other \cite{CPhysRep}. 
 From the present results  we
see, however, that this is not the case. Upon propagation, the beam tends to
reach the on-axial value of the intensity which {\it maximizes} the index of
refraction at the beam axis. In other words, not the nonlinear refractive
index, but its variation should be zero: 
\begin{equation}
\partial_I n(I)\vert_{I_{\rm sat}} =0.\label{Isat}
\end{equation}
This new condition is general, independent of the medium or material
and represents one of the predictions of this letter, which should
serve as a basis for future calculations. Note that such a general
mathematical condition would have been impossible to obtain 
on the basis of numerical simulations.

We now apply our theoretical scheme to study the concrete problem of
femtosecond laser pulse propagation in air, which is relevant 
 due to a large number of 
applications and whose description is still a subject of discussion 
(see, e.~g.\ Refs.~\cite{CPhysRep,BergeRPP} and
Refs.~therein). The nonlinear refractive index of air is
 taken in the following widely used form 
\begin{eqnarray}
n=n_2[I+f(I)]+n_4[I^2+f(I^2)]-{\rho (I)\over 2\rho_c},\label{n_air}
\end{eqnarray}
where the first term describes the Kerr response involving a delayed (Raman)
contribution $f(I)=\tau_K^{-1}\int_{-\infty}^te^{-(t-t')/\tau_K}I(t')dt'$. 
$n_2$ and $\tau_K$ are known to be equal to
$3.2\times 10^{-19}{\rm\ cm/W^2}$ and $70{\rm\ fs}$ \cite{BergeTera}, respectively. 
 $n_4\equiv\chi^{(5)}/2n_0$, where $\chi^{(5)}$ is the fifth order nonlinear susceptibility its exact value is a subject of some controversy \cite{BergeRPP,Berge_Chi5}. 
The most accepted estimates lie around $\sim 10^{-32}{\rm\ cm^4/W^2}$
\cite{CPhysRep}. In the last term of Eq.~(\ref{n_air}), $\rho(I)$ refers to the density of free electrons
and $\rho_c$ denotes the critical density above which the plasma
becomes opaque.  A rough estimate yields $\rho(I)\sim\sigma_K I^K \rho_{\rm
at}t_p$, where $\rho_{\rm at}$ is the atom
density $\rho_{\rm at}=2\times 10^{19}{\rm\ cm^{-3}}$ and $t_p$  the
pulse duration. $K=8$ for the MPI with a pulse of $800{\rm\ nm}$, and  
$\sigma_8=3.7\times 10^{-96}{\rm\ cm^{16}/W^8/s}$ \cite{CPhysRep}.

A numerical solution of the NLSE for the pulse with $P_{\rm in}\sim 0.08{\rm TW}$ and $w_0=3 {\rm mm}$ with $n(I)$ given by Eq.~(\ref{n_air}) with $n_4=0$ gives $z_{\rm sf,f}=128.18 {\rm cm}$  Ref.~\cite{Tzor}. Note that if $n(I)=n_2I$, the self-focusing distance is given by
the Kovalev formula \cite{Prep,ShirKovPhysRep} $z_{\rm
sf}=w_0/(2\sqrt{n_2I_0})$, which is exact under the geometrical optics
approximation and for initially Gaussian beam shape. 
For the experimental conditions of Ref.~\cite{Tzor} this yields 
 $z_{\rm sf,f}=127.66 {\rm cm}$. This confirms our initial statement 
 that for many  experiments  the diffraction
(and in this case also the plasma defocusing) can be neglected. 

Recently reported experimental results on air clearly indicate that $z_{\rm
sf}$ scales as $1/\sqrt{P_{\rm in}}$ for a relatively low initial
pulse power $P_{\rm in}$. However, for powers above $500{\rm\ GW}$
($I_0 \sim 5 \times 10^{12}{\rm W/cm^2}$) a qualitative change is observed
and $z_{\rm sf}$ depends on the power as $\sim 1/P_{\rm
in}$ \cite{Fibich}.  This behavior was attributed by authors to noise
effects in the beam.  On the other hand one can notice
that the value of $500{\rm\ GW}$ in the experiment of
Ref.~\cite{Fibich} corresponds to the case where the terms
$n_2I_0=1.7\times 10^{-6}$ and $n_4I_0^2\sim 3.1\times 10^{-7}$ in the
refractive index (\ref{n_air}) become comparable. Therefore, we
believe that the change in the power dependence is mainly driven by the
contribution of the highly nonlinear terms (fifth order
susceptibility).  Moreover, from the analysis of the experiment of
Ref.~\cite{Fibich} within our theoretical scheme we can draw important
conclusions regarding the form of the nonlinear refractive index.

\begin{figure}
\includegraphics[angle=0,width=10cm]{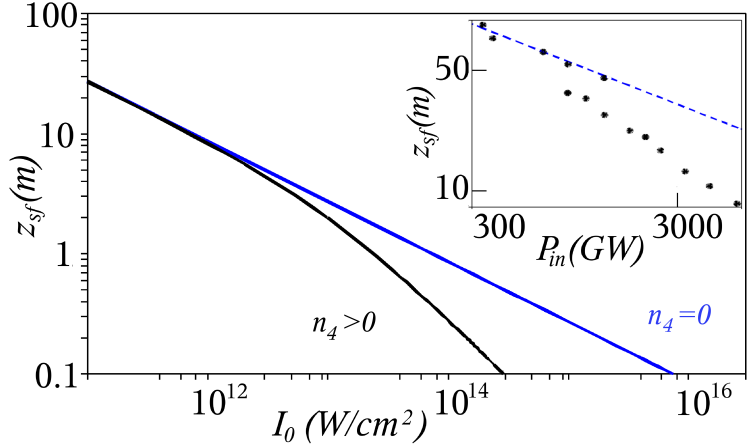}
\caption{\label{z(bI)} (Color online.)  Dependence of the
self-focusing position on the beam intensity. The blue curve refers to
the dependence $1/\sqrt{I}$ (obtained for low intensities), whereas
the black curve shows the deviation  for high intensities due to
influence of the fifth-order nonlinearity $n_4=10^{-32}{\rm
cm^4/W^2}.$ Inset: Experimental points for $z_{\rm fs}(P_{\rm in})$
from Ref.~\cite{Fibich}.}
\end{figure}

Using the value of $I_0$ given in Ref.~\cite{Fibich}, and assuming  
$n_4 \sim \mp 10^{-32}{\rm\ cm^4/W^2}$ \cite{CPhysRep}, we get $\beta\simeq
\pm 0.35$.  Comparing this value with $\beta_c\sim 0.175$, we see that, if $n_4 < 0$, 
 then no self-focusing would be observed. Since self-focusing is
 indeed observed, we conclude that $n_4>0$. The behavior of $z_{\rm
 sf}$ as a function of the initial beam intensity obtained from our
 theory is shown in Fig.~\ref{z(bI)} and is qualitatively compared to
 the measured $z_{\rm sf}$. Note that one could use our results to find an accurate 
 value of $n_4$ by performing a 
  similar experiment with a controlled initial Gaussian  beam profile (i.e., without 
 noise) and fitting the  measured curve $z_{\rm sf}(I_0)$ to the
  solution 
 of Eq.~(\ref{sol_(1-bn)}).

\begin{figure}
\includegraphics[angle=90,width=10cm]{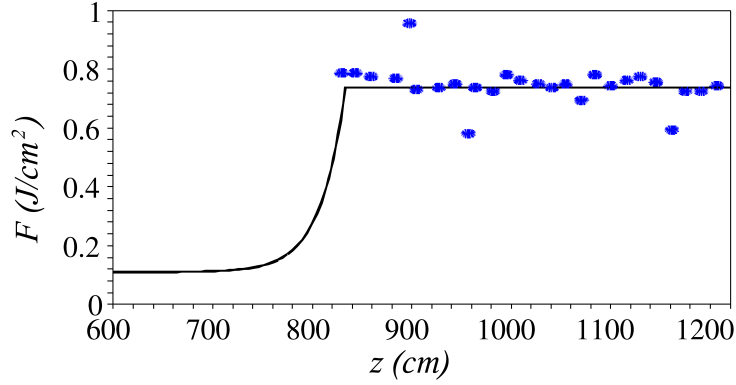}
\caption{\label{CompFig} Filamentation in air. On-axial fluence
versus the propagation distance. Circles are the experimental results from
Ref.~\cite{Ting}. The black curve represents our analytical solution.}
\end{figure}

Based on the results of this work it is clear that one is able to 
  determine experimentally the nonlinear optical constants by
 measuring $I_{\rm sat}$. As we discussed above, the last term 
in Eq.~(\ref{n_air}) does not change $z_{\rm sf}$
significantly. However, its value is crucial for $I_{\rm sat}$. In
Ref.~\cite{Ting} $z_{\rm sf}$ and $I_{\rm sat}$ for air were
measured. In the experimental setup a collimated beam with FWHM diameter of 
$d\sim 4{\rm\ mm}$ ($w_0=d/\sqrt{2\ln 2}$), a pulse duration (FWHM) of 
$450{\rm\ fs}$ and a pulse energy $E_{\rm in} \sim 20{\rm\ mJ}$ was  
applied. We took these values and used them to calculate the on-axial intensity 
distribution from our analytic equations. The comparison between our theory and 
experimental results of Ref.~\cite{Ting} is shown
in Fig.~\ref{CompFig}.  The temporal pulse compression due to MPI is 
estimated along the similar lines with Ref.~\cite{BergeTera}. For the
refractive index given by Eq. (\ref{n_air}), we calculate $I_{\rm
sat}$ on the basis of Eq.~(\ref{Isat}). The value of $n_4$ chosen
in a way to provide a satisfactory fit of experimental date for both
$z_{\rm sf}$ and $I_{\rm sat}$ is equal to $8 \times 10^{-32}{\rm\
cm^4/W^2}$. It is necessary to point out that for a more accurate 
 determination additional
experimental studies are required.

In Ref.~\cite{Ting} an additional experiment was reported using a shorter pulse of
 duration (FWHM) $50{\rm\ fs}$ and energy $19{\rm\ mJ}$. For the
same model input-parameters, we obtain: $I_{\rm sat}=1.26{\rm\ W/cm^2}$ and
the fluence $F=0.67{\rm\ J/cm^2}$ which are in a good agreement with
the experimental results $I_{\rm sat}=1.3{\rm\ W/cm^2}$ and
$F\sim0.6{\rm\ J/cm^2}$, correspondingly.

In Table~\ref{CompTable} we demonstrate the influence of high order
nonlinearities on the self-focusing position. The Marburger Formula
(MF) reads $z_{\rm sf}=0.367z_0\left((\sqrt{P_{\rm in}/P_{\rm
cr}}-0.852)^2-0.0219\right)^{-1/2}$; here $z_0=\pi w_0^2/\lambda \sim 45{\rm m}$,
$P_{\rm cr}=\lambda^2/2\pi n_2\simeq 3.3{\rm GW}$. The ``effective
critical power`` mentioned  in the table means a replacement $P_{\rm cr}\to P'_{\rm cr}$,
where the relation $P'_{\rm cr}=2P_{\rm cr}$ was used 
\cite{Tzor}.  From the Table \ref{CompTable} one can see that other 
analytical approaches which do not account for high order nonlinearities
 cannot provide a satisfactory prediction of $z_{\rm sf}$ in modern high-intensity experiments.

\begin{table}
\caption{\label{CompTable} Comparison of the predictions for $z_{\rm sf}$ in air calculated using different theoretical approaches
  and compared with the experimental data of Ref.~\cite{Ting}.}
\begin{tabular}{|l|r|}\hline
Experiment & $\sim 8{\rm\ m}$ \hspace{1cm}\\ Marburger formula (MF)
\cite{Marburger} & $6.1{\rm\ m}$ \\ MF with an effective critical
power \cite{CPhysRep} & $9.9{\rm\ m}$ 

\\ Kovalev formula ($n_4=0$)
\cite{Prep} & $9{\rm\ m}$ \\ 

Eq.~(\ref{sol_(1-bn)}) for $n_4=8 \times 10^{-32}{\rm cm^4/W^2}$                 & $8.3{\rm\ m}$                \\
\hline
\end{tabular}
\end{table}

Finally,  and for the sake of completeness, we present  results
obtained by applying our theory to other types of nonlinearities in order to predict 
 the behavior of self-focusing for other cases  of 
  current physical interest \cite{BergePhysRep}. 
 If the nonlinear refractive index has the form 
 $n(I)=a I/(1+\beta I)$, the on-axial intensity distribution is given by the expression $$
[(1+\beta I)^3-a 3\beta I^2z^2]^{1/3}-1=\beta.$$
For $n(I)=1-e^{-\beta I/a}$, the on-axial intensity is given by $$e^{\beta I}-a z^2\beta^2I^2=e^\beta,$$ and 
for a polynomial form $n(I)=I^k$ ($k\neq 2$) by $$I^2[I^{-k}+k(k-2)a z^2]=1.$$ No intensity saturation has been observed.

The singularities in solutions of equations above correspond
to the $z_{\rm sf}$ for the given $n(I)$. It should be noted that
the approach suggested in this letter can be generalized to arbitrary $n(I)$. In general, the
solutions can be found semi-analytically by interpolation of the integrals
in Eqs.~(\ref{sol}).

Summarizing, exact solutions for the self-focusing length
$z_{\rm sf}$ and the filament intensity $I_{\rm sat}$ for several different forms of the nonlinear refractive index in the
framework of geometrical optics were obtained.  Depending on the
experimental conditions, these solutions can be very accurate,
describe the essential physics of the problem and explain different
independent measurements.  
 The analytical expressions obtained for the dependence $I = I(z)$ constitute a clear improvement 
 with respect to the empirical  Marburger formula.

This work has been supported by the DFG
through SPP 1134.


\begin{thebibliography}{99}

\bibitem{Diels} J.-C. Diels and W. Rudolph, {\it Ultrashort Laser Pulse Phenomena} (Elsevier, Amsterdam,  2006). 

\bibitem{Englert}  L. Englert {\it et al.,} Opt. Express {\bf 15} 17855 (2007). 

\bibitem{Woeste} K. Stelmaszczyk {\it et al.,} Appl. Phys. Lett., {\bf 85,} 3977 (2004).


\bibitem{Chem} J. Zhang {\it et al.,}�J. Am. Chem. Soc. {\bf 128,} 406 (2006).

\bibitem{MacrMol} C. Degoulet {\it et al.,} Macromolecules {\bf 34,} 2667  (2001).

\bibitem{Med} G. Paltauf and P. E. Dyer, Chem. Rev. {\bf 103,} 487 (2003).

\bibitem{CPhysRep} A. Couairon and A. Mysyrowicz, Phys. Rep. {\bf 441,} 47 (2007). 

\bibitem{BergeRPP} L. Berg\'e {\it et al.}, Rep. Prog. Phys. {\bf 70,} 1633 (2007).

\bibitem{Marburger} J. H. Marburger, Prog. Quant. Electron. {\bf 4,} 35 (1975).  Under the
geometrical optics approximation, and for the same form of refractive
index, exact analytical expressions for $z_{\rm sf}$ 
 have been obtained in Refs. 
\cite{Akhmanov} and \cite{Prep,ShirKovPhysRep}.

\bibitem{Akhmanov} S. A. Akhmanov {\it et al.,}  Sov. Phys, Usp. {\bf 10,} 609 (1968).

\bibitem{Prep} V. F. Kovalev, JINR Prep. No. P5-96-477, Dubna (1996).  

\bibitem{ShirKovPhysRep} D. V. Shirkov and V. F. Kovalev, Phys. Rep. {\bf 352,} 219 (2001).

\bibitem{BergePhysRep} L. Berg\'e, Phys. Rep. {\bf 303,} 259 (1998).

\bibitem{Fibich} G. Fibich {\it et al.}, Opt. Express {\bf 13,} 5897 (2005).

\bibitem{noncollim} Note that for focused beams the self-focussing position has to be corrected to $z_{\rm sf,f}$ which satisfies  
$ 1/z_{\rm fs,f}=1/z_{\rm sf} + 1/f,$ 
where $f$ is the focal length of the lens (see Ref. \cite{CPhysRep}).

\bibitem{Bocharov} A. V. Bocharov, {\it Symmetries and Conservation Laws for Differential Equations of Mathematical Physics} (AMS
, 1999).

\bibitem{Kuramoto} Y. Kuramoto, {\it Chemical oscillations, waves, and turbulence}, (Springer, Berlin, 1984).

\bibitem{GKZ} A. N. Gorban {\it et al.,}  Phys. Rep. {\bf 396,} 197 (2004).



\bibitem{CPRA} A. Couairon, Phys. Rev. A {\bf 68,} 015801 (2003).

\bibitem{Berge_Chi5} A. Vin\c{c}otte and L. Berg\'e, Phys. Rev. A {\bf 70,} 061802(R) (2004).

\bibitem{Note} Previosly, an approximate analytical expression for $z_{\rm fs}$ for arbitrary $n(I)$ was obtained in Ref. \cite{TatGar}, its accuracy was $\sim a^2 z^3.$

\bibitem{TatGar} L. L. Tatarinova and M. E. Garcia, Phys. Rev. A {\bf 76,} 043824 (2007).

\bibitem{BergeTera} L. Berg\'e, {\it et al.}, Phys. Rev. Lett. {\bf 92,} 225002 (2004).
 
\bibitem{Tzor} S. Tzortzakis {\it et al.}, Phys. Rev. Lett. {\bf 86,}  5470 (2001).

 
\bibitem{Ting} A. Ting {\it et al.}, Phys. Plasmas {\bf 12,} 056705 (2005). 


\end{thebibliography}
\end{document}